**ORIGINAL ARTICLE**  OPEN ACCESS

# Artificial Intelligence and Nuclear Weapons Proliferation: The Technological Arms Race for (In)visibility

David M. Allison[1]  |  Stephen Herzog[2,3,4]

[1]Political Science Department and Yale Nuclear Security Program, Yale University, New Haven, Connecticut, USA | [2]Center for Security Studies, ETH Zurich, Zurich, Switzerland | [3]James Martin Center for Nonproliferation Studies, Middlebury Institute of International Studies at Monterey, Monterey, California, USA | [4]Project on Managing the Atom, Belfer Center for Science and International Affairs, Harvard Kennedy School of Government, Harvard University, Cambridge, Massachusetts, USA

**Correspondence:** Stephen Herzog (stephen.herzog@sipo.gess.ethz.ch)

**ABSTRACT**

A robust nonproliferation regime has contained the spread of nuclear weapons to just nine states. Yet, emerging and disruptive technologies are reshaping the landscape of nuclear risks, presenting a critical juncture for decision makers. This article lays out the contours of an overlooked but intensifying technological arms race for nuclear (in)visibility, driven by the interplay between proliferation-enabling technologies (PETs) and detection-enhancing technologies (DETs). We argue that the strategic pattern of proliferation will be increasingly shaped by the innovation pace in these domains. Artificial intelligence (AI) introduces unprecedented complexity to this equation, as its rapid scaling and knowledge substitution capabilities accelerate PET development and challenge traditional monitoring and verification methods. To analyze this dynamic, we develop a formal model centered on a Relative Advantage Index (RAI), quantifying the shifting balance between PETs and DETs. Our model explores how asymmetric technological advancement, particularly logistic AI-driven PET growth versus stepwise DET improvements, expands the band of uncertainty surrounding proliferation detectability. Through replicable scenario-based simulations, we evaluate the impact of varying PET growth rates and DET investment strategies on cumulative nuclear breakout risk. We identify a strategic fork ahead, where detection may no longer suffice without broader PET governance. Governments and international organizations should accordingly invest in policies and tools agile enough to keep pace with tomorrow's technology.

## 1 | Introduction

Will the age of readily detectable nuclear weapons proliferation draw to a close? Recent Israeli and US airstrikes on Iranian nuclear facilities underscore the urgency of detecting potential covert developments before maturation. Less well-known to the public and politicians is the growing role of emerging technologies in reshaping the nuclear risk landscape. Tools and expertise once exclusive to advanced nuclear-armed states—high-performance computing (HPC) simulations, precision manufacturing, and specialized nuclear engineering—are increasingly accessible. Additive manufacturing, affordable microprocessors, and automated design software lower barriers to weapon development (Christopher 2015; Vaynman & Volpe 2023; Volpe 2019). Most strikingly, rapid advances in artificial intelligence (AI), particularly large language models (LLMs), may soon substitute machine-generated expertise for the skilled personnel traditionally required to design and conceal the bomb (Kaplow & Musto 2025; Meier 2024; Pan & Cheng 2025; Reinhold & Schörnig 2022). While much scholarship on AI and nuclear risks focuses on command-and-control systems, strategic stability, and escalation (Allison 2025; Boulanin 2019; Cox &

Williams 2021; Geist 2023; Horowitz & Lin-Greenberg 2022; Johnson 2023; Lin 2025; Schwartz & Horowitz 2025), relatively little systematic academic research addresses the potential role of LLMs in enabling nuclear arms proliferation.

These capabilities are driving an accelerating—but often unseen—arms race centered on nuclear (in)visibility. The contest pits innovations enabling clandestine nuclear activities against those enhancing detection. Institutions like the International Atomic Energy Agency (IAEA), supported by states' national technical means (NTM, including satellite reconnaissance and signals intelligence), kept odds of undetected proliferation low for much of the post–Cold War period (Roehrlich 2022). Their success stemmed from the difficulty of acquiring fissile material and the visibility of large-scale nuclear infrastructure. But that balance may be shifting. AI systems without robust guardrails, additive manufacturing pipelines outside export controls, and commodified dual-use components all threaten nonproliferation (Montgomery 2020; Nelson 2020). Simultaneously, safeguards and monitoring technologies are evolving: AI-enhanced satellite imagery analysis, algorithmic anomaly detection in trade data, and autonomous inspection platforms may offset proliferators' gains (Baldus 2022; Kaspersen & King 2019; Lee et al. 2025; Onderco & Zutt 2021).

This article introduces a framework for evaluating the balance between proliferation-enabling technologies (PETs) and detection-enhancing technologies (DETs). If PETs improve faster than DETs, covert nuclear weapon acquisition becomes more likely. If DETs lead, detection remains feasible. Rapid AI scaling introduces profound uncertainty. Whereas LLMs in 2023 demonstrated performance on certain standardized tests comparable to undergraduates (Frieder et al. 2023), 2025 models are showing competence at specialized scientific and engineering skills once considered exclusive to human experts. While this expertise is often narrow and models can lack the reliability of human researchers, the pace of advancement continues to surprise observers and underscores potential for rapid changes in proliferation-relevant capabilities.[1] In worst-case scenarios, frontier systems may rival entire weapon design teams within just a few years, especially in compressing tacit knowledge (Budding 2025; Lu 2025). Whether detection capabilities can comparably scale remains unknown, given cumbersome bureaucratic and institutional constraints (Marchant 2011; Stewart & Mayhew 2019).

This study first reviews traditional tools of nonproliferation monitoring and evasion strategies states have pursued. We then develop a formal model centered on a Relative Advantage Index (RAI), quantifying gaps between proliferators' concealment capabilities and detectors' observation abilities. The index is not intended as a predictive tool, but rather as a heuristic framework to structure analysis of this technological arms race and clarify the stakes of different policy choices. We simulate how potentially asymmetric trajectories—like AI-driven PET growth versus stepwise DET upgrades—could create a widening band of uncertainty around nuclear weapons proliferation risk and detectability. We evaluate multiple scenarios: from "Limited AI," where PET progress plateaus, to "Transformative AI," where rapid, near-exponential PET growth removes key proliferation constraints. Finally, we discuss policy implications, including where early detection investments can still shift outcomes and where risk governance and demand-side interventions become pivotal.

## 2 | Nuclear Proliferation and Nonproliferation

Nonproliferation has historically relied on institutional agreements—principally the Treaty on the Non-Proliferation of Nuclear Weapons (NPT)—and technical monitoring (Coe & Vaynman 2015; Gibbons & Herzog 2022; Hunt 2022; Miller 2018). Understanding this framework is crucial to appreciate the transformative impact of new technologies. NPT non-nuclear-weapon states commit to not acquiring the bomb and accept IAEA safeguards inspections at civilian nuclear facilities. Monitoring and verification focus on declared sites and materials, employing periodic on-site inspections, seals, cameras, and facility operating record reviews (Roehrlich 2022; Zubair et al. 2024). These measures are augmented by advanced states' NTM data that are often shared with international authorities (Carnegie & Carson 2019). During the late Cold War and into the 1990s, this toolkit proved relatively robust at detecting violations.

Limitations became evident after the discovery of Iraq's extensive clandestine nuclear program in 1991 (Findlay 2022). This program, largely missed by the safeguards regime, exposed proliferation loopholes like undeclared sites and foreign procurement. North Korea's early 1990s activities similarly highlighted shortcomings in verifying completeness and accuracy of state declarations. These experiences catalyzed a major safeguards overhaul, notably the Additional Protocol granting the IAEA broader inspection authority in consenting states (Gibbons & Robinson 2021). Later, revelations about the A.Q. Khan illicit procurement network and Iran's covert activities indicated a need to move beyond passive inspections to include open-source intelligence and advanced analytics (Salisbury 2014).

Nuclear proliferators adopt strategies based on the international monitoring environment and their own technical capabilities. As Narang (2022) explains, there are distinct approaches, including building latent capabilities while deferring full weaponization ("hedging"), secretly pursuing weapons ("hiding"), or rapidly racing toward a bomb ("sprinting"). Hedging builds advanced technical capability under legal cover of a civilian program, avoiding weaponization decisions when detection capabilities are strong. Hiding aims for a *fait accompli* arsenal by maximizing secrecy, covert procurement, and deception when international detection capabilities may be circumvented. Sprinting mobilizes resources quickly and openly when detection fears are secondary to weaponization imperatives. Strategies are fluid: States may shift from hedging to hiding if detection pressure weakens, or from hiding to sprinting if they fear being exposed (Narang 2016).

Critically, these strategies hinge on the detection environment. Weaponization—the final stages of assembling a deliverable nuclear weapon—is difficult to observe. Unlike uranium enrichment or plutonium reprocessing, which produce detectable emissions or require large observable infrastructure, weaponization can occur in small laboratories or dispersed facilities (Allison et al. 2020; Herzog 2020). As such, monitoring has long focused on fissile material production as a constraint and a warning sign (Herzog 2025). Detecting material diversion from civilian nuclear programs and locating clandestine facilities are thus cornerstones of the nonproliferation regime (Glaser et al. 2008).

But a new challenge lies in asymmetric evolution of PETs and DETs, which have different rhythms and motivating incentives. PETs span tools that simplify bomb pathways: advances in hardware, software, and even cognitive substitutes for human expertise. Additive manufacturing (3D printing) of maraging steel for centrifuge rotors and computer numerical controlled (CNC) machining for explosive lenses may reduce needs for conspicuous industrial facilities (Christopher 2015). These tools are maturing fast and diffusing widely, shrinking the potential footprint of weapons programs.

Perhaps the most radical new PET is AI. Models as early as GPT-4 showed potential utility in weapons of mass destruction proliferation (Brent & McKelvey Jr. 2025; Shoker et al. 2023; Stendall et al. 2024). Critically, AI may reduce tacit knowledge barriers—the "how" of nuclear design and fabrication—and can be coupled with other PETs. LLMs trained on open scientific literature, simulation data, and technical manuals may walk less experienced personnel through weaponization steps (OpenAI 2023, 52).

The lack of commercial incentives for proliferation-enabling algorithms is unlikely to slow this process. Task-specific AI tends to follow market demand: If no industry needs a reactor core optimization model, capabilities are unlikely to emerge. By contrast, LLMs are general purpose. Commercial incentives accelerate PETs by pushing frontier models to master the scientific stack. Models excel at *all* mathematics, coding, and scientific reasoning tasks because doing so broadens downstream revenue (Berti et al. 2025). A model for automating spreadsheets or drug discovery inherits latent competence in explosives geometry. These tools will likely benefit both aspiring proliferators and established nuclear-armed states. For the latter, they could enable breakthroughs like pure-fusion warhead development, dispensing with today's thermonuclear designs that rely on a fission primary to ignite a fusion secondary.

Moreover, AI's implications extend beyond warhead design to the fuel cycle itself. LLMs may guide decisions about centrifuge cascade design, reactor modeling, and spent fuel reprocessing (Almeldein et al. 2025). Combined with additive manufacturing, this may allow the manufacture of parts for fissile material production, subverting strategic trade controls (Christopher 2015; Volpe 2019).

While little systematic scholarship exists at the AI–nuclear proliferation nexus, there has been some nonproliferation optimism (Bajema 2024; Pan & Cheng 2025). For example, Montgomery (2025) suggests states may struggle to adopt and integrate emerging technologies into nuclear programs. This argument draws a line between AI and historical cases of hardware-centric technologies, where complexity, tacit knowledge requirements, and weak institutions slowed nuclear proliferation. Yet, the ubiquity of commercial, general-purpose AI means states are likely to already have skilled personnel. Consider, by analogy, how Ukraine's drone-racing community became the nucleus of its military drone operations (Kunertova 2023). Prospective AI–nuclear proliferation pathways differ from past cases by diffusing and evolving rapidly, automating tacit knowledge transfer, and potentially integrating with additive manufacturing to circumvent supply chains. As Debs & Monteiro (2016) argue, proliferation hinges on both demand-side willingness and supply-side opportunity. AI reduces opportunity costs by lowering technical and cognitive barriers. States previously deterred by complexity may come to view a bomb as feasible.

HPC and advanced modeling software will further transform research and development. Calculations and design validation once requiring extensive human effort or full-scale testing can increasingly be simulated (Kim et al. 2020). Many relevant tools are open source or commercially available, helping proliferators sidestep more observable testing activities. In parallel, digital knowledge networks, including online forums and dark web platforms, accelerate the diffusion of nuclear know-how.

Collectively, PETs—especially AI, modeling software, and manufacturing technology—are advancing rapidly. Their growth often follows a logistic or S-curve pattern, characterized by initial near-exponential advancement that eventually slows as technology matures. This growth is driven by commercial incentives—largely indifferent to nonproliferation concerns (Herzog & Kunertova 2024). These trends suggest that would-be proliferators will have a continuously improving ability to indigenize the nuclear fuel cycle and pursue the bomb.

DETs are also evolving (Vaynman 2021). Satellite and remote sensing innovations have expanded data coverage and resolution. Commercial low-earth orbit (LEO) satellites provide daily high-resolution imagery; new sensors, such as thermal infrared satellites, can reveal heat signatures from reactors or enrichment facilities (Fetter & Sankaran 2025). AI techniques are increasingly used to automate analysis of imagery (Kaspersen & King 2019). Combined thermal–AI pipelines have shown early promise, although autonomous detection remains difficult because of deliberate obfuscation of nuclear signatures (Park et al. 2024).

AI-driven data fusion also integrates disparate monitoring signals. The Preparatory Commission for the Comprehensive Nuclear-Test-Ban Treaty Organization (CTBTO) employs machine learning to distinguish between earthquakes and nuclear explosive tests, combining seismic, infrasound, hydroacoustic, and radionuclide data (Kalinowski et al. 2023). National intelligence agencies use pattern recognition software to identify illicit procurement among trade flows, financial logs, and shipping data (Lee et al. 2025; Salisbury 2014). Meanwhile, the IAEA is applying AI—from machine learning to digital twins—to enhance remote monitoring and inspections (Ritter et al. 2022). Automated image and video analysis can flag anomalies, and drones can perform tasks in hazardous or difficult-to-reach areas. Societal verification using open-source materials, commercial satellite imagery, and scientific sensors for geophysical monitoring adds another layer to the DET landscape (Al-Sayed 2022).

But DETs are historically deployed in stepwise bursts. The US CORONA satellite program, triggered by Cold War anxieties, revolutionized imagery (Albright et al. 2018; Bateman 2020). The discovery of Iraq's nuclear weapons program led to the Model

Additional Protocol (Findlay 2022; Gibbons & Robinson 2021). The A.Q. Khan network and Libya's proliferation disclosures sparked innovative trade, shipping, and network data fusion (Salisbury 2014). Yet, each leap came as a result of crises.

This points to a worsening pacing problem due to rapidly emerging technologies (Marchant 2011). While many innovations are easily assimilated by existing legal and regulatory structures, dual-use technologies like AI can create governance gaps as institutional adaptation struggles to keep up with technological change (Canfil 2025b). Verification agencies struggle not just to regulate frontier PETs, but also to adopt DETs. Promising algorithms may sit idle for years, delayed by sovereignty concerns, commercial secrecy, validation protocols, or bureaucratic inertia. Detector development remains bespoke and underfunded (International Atomic Energy Agency 2022), while proliferators benefit from commercial incentives and access to general-purpose technology. In the long run, this asymmetry may create a temporal advantage for proliferators seeking to evade detection. Our contribution highlights the need to quickly address this dual lag and prevent institutions from falling behind in regulating and deploying technology.

## 3 | The Relative Advantage Index

To this end, we develop a formal model of the race between PETs and DETs. The purpose of this model is not to predict the future, but to impose analytical discipline and contextualize the stakes at play. Formalization allows us to systematically explore how different technological trajectories alter nuclear proliferation risks and assess the potential impact of policy interventions under varying conditions.

To analyze PET and DET evolution, we introduce the RAI. The RAI captures the shifting proliferation and detection balance, illustrating how uncertainty widens over time. At time $t$, RAI is defined as the difference between a proliferator's capability to evade detection, $P(t)$, and the international community's detection capability, $D(t)$: $RAI(t) = P(t) − D(t)$. Although $P(t)$ and $D(t)$ are abstract indices, their initial values and growth rates reflect plausible relative starting points and change dynamics. We set $D(0)$ 50 to represent current baseline global detection capability and $P(0)$ 20 such that the starting RAI is −30. These values for $P(t)$ and $D(t)$ reflect current consensus that detection capabilities hold a meaningful advantage over proliferation-enabling capabilities (Carnegie & Carson 2019; Roehrlich 2022). Crucially, our model focuses not on absolute values but on the gap between the two indices and their subsequent rates of change.[2] When RAI is positive, proliferators are more likely to succeed undetected; when negative, detection dominates.

RAI shifts can influence state behavior. When proliferators perceive $P(t)$ outpacing $D(t)$, a positive RAI, they may pursue hiding, banking on new tools to evade detection. Others might hedge, betting that a shift into weaponization will go unnoticed (Narang 2022). Higher RAIs reduce perceived risks of clandestine activity. Conversely, a negative or low RAI reflects increased detection risks, encouraging caution. States may accordingly hedge within legal bounds, developing lab- or pilot-scale safeguarded enrichment or reprocessing facilities rather than attempting diversion or weaponization (Fuhrmann 2024; Hymans 2006). In detection-favorable environments, hiding becomes less viable and sprinting more perilous (Debs & Monteiro 2016). When RAI shifts mid-program, states react. A sudden DET leap turning the RAI negative might force a hiding state to pause, roll back, or sprint. A strong detection regime deters proliferation, but if successfully challenged, that deterrent may erode, triggering wider breakout behavior.

Two key AI characteristics dramatically reshape RAI. First, LLMs are scaling extremely rapidly. A likely trajectory for AI-driven capabilities is a logistic curve, featuring an initial growth period that doubles capabilities within months or years, followed by a slowdown as technology approaches inherent limits.[3] This dynamic, driven by commercial innovation, informs our modeling of PETs as a smooth logistic curve. In contrast, DETs rely on governmental and multilateral processes, tending to improve in sporadic jumps reacting to crises. This asymmetry complicates forecasting, as gains can quickly compound into major disparities. Second, AI exacerbates the challenge of characterizing uncertainty in the proliferation landscape. The field of risk analysis distinguishes between aleatory uncertainty, which represents a system's inherent randomness, and epistemic uncertainty, which stems from a lack of knowledge about the system's true properties or functional forms (Aven 2010). While some elements of (non)proliferation involve aleatory uncertainty—like the chance of success of a specific intelligence operation—the dominant challenge posed by emerging technologies is a profound increase in epistemic uncertainty. Fast, unpredictable AI development makes it difficult to specify, let alone place confidence in, key parameters governing both nuclear weapons proliferation and detection. This uncertainty complicates efforts to forecast the future strategic balance. Figure 1 illustrates a possible change in PETs and DETs over a decade. Shaded areas around each curve represent uncertainty associated with these respective capability trajectories.

PET scaling and deep epistemic uncertainty undermine steady-state planning as predictive confidence in $P(t)$ and $D(t)$ balance naturally diminishes over longer time horizons. Today, detectors likely hold the advantage. But in a few years, open-source AI could potentially enable in silico weapon design and covert enrichment. Conversely, parallel AI-driven detection might yield massively enhanced surveillance, though such capabilities will always be subject to fundamental physical and practical limits. Both scenarios are plausible, and the space between them is expanding. This epistemic uncertainty poses major challenges for risk assessment and policy planning. Analysts lack the data needed to confidently specify probability distributions for key variables, a common feature of forecasting complex and non-stationary sociotechnical systems. While subjective judgments are possible, and indeed necessary for modeling (Cooke 1991), low confidence in these judgments motivates our scenario-based approach exploring multiple plausible futures instead of generating a single forecast. Shaded areas in Figure 2 illustrate RAI uncertainty.

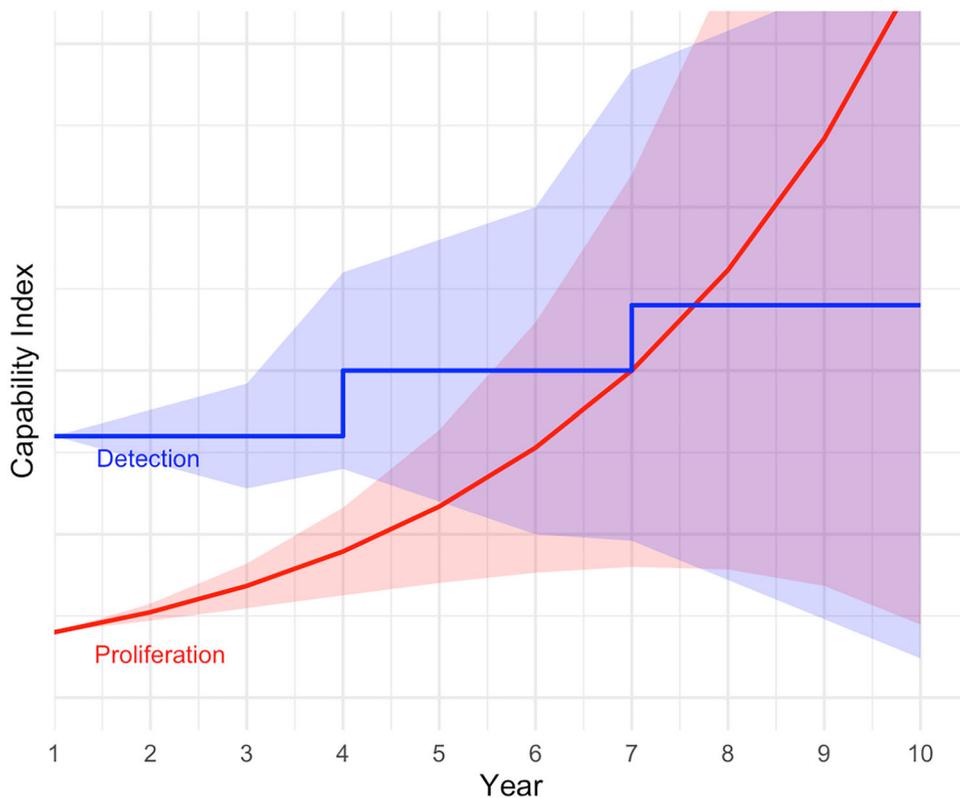

**FIGURE 1** | Theorized logistic PET growth versus stepwise DET improvements.

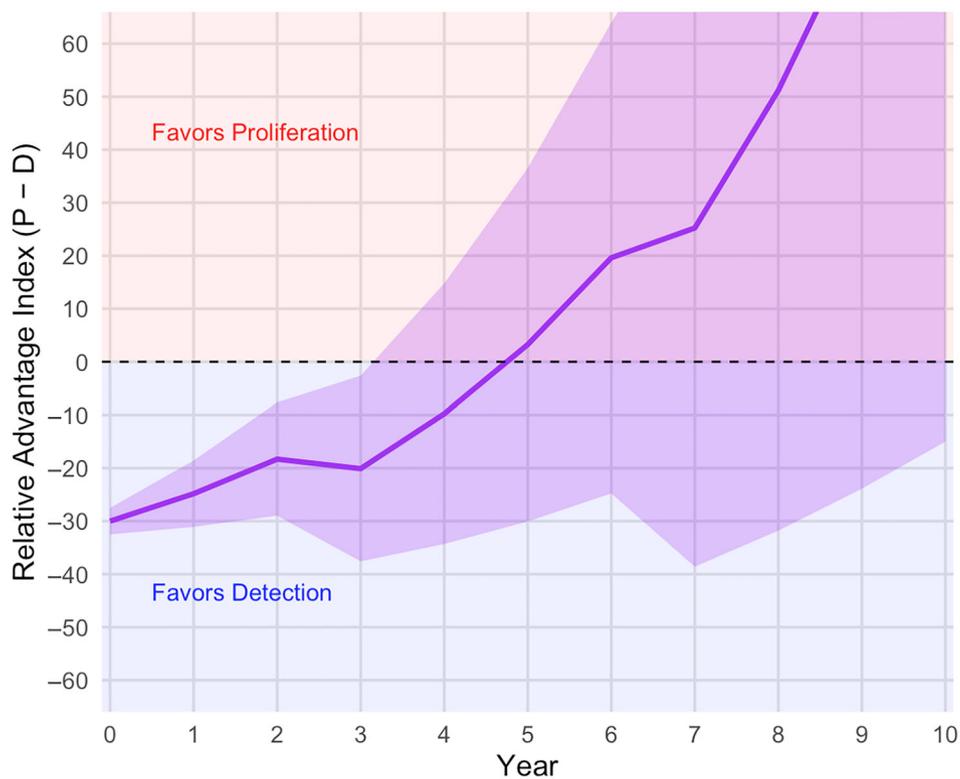

**FIGURE 2** | Relative Advantage Index and widening uncertainty band.

## 3.1 | Modeling Proliferation Risk

We extend the RAI to model proliferation risks within planning horizon $T$. Cumulative probability that a clandestine program evades discovery within planning horizon $T$ is

$$R(T) = \int_0^T \lambda(t)\left[1 - \Pr(D_t)\right] dt, \quad (1)$$

where $\lambda(t)$ is the rate of clandestine breakout attempts and

$$\Pr(D_t) = \frac{1}{1 + \exp[-(\kappa\, \text{RAI}(t) + \theta)]} \quad (2)$$

maps the instantaneous RAI gap to detection probability. We set $\kappa = 0.4$ to match IAEA safeguards modeling standards, which target a $\approx 90\%$ detection probability for significant material diversions (International Atomic Energy Agency 2002, 24), implying a $\approx 10\%$ false-negative rate for undeclared enrichment.[4] We set $\theta = 0$ so that $\Pr(D_t) = 0.5$ whenever $RAI(t) = 0$.[5] The elasticity of risk concerning RAI is

$$\varepsilon_R = \frac{\partial R/\partial \text{RAI}}{R/\text{RAI}}, \quad (3)$$

providing a concise measure of how aggressively policymakers must act to check breakout risks.

## 3.2 | Capping PET Effects

Even with superintelligent design aids, throughput and material constraints bound capabilities.[6] We impose a logistic ceiling,

$$P(t) = \frac{P_{\max}}{1 + \left(\frac{P_{\max}}{P_0} - 1\right)e^{-gt}}, \qquad P_{\max} = 120, \quad (4)$$

reducing to the exponential form for $P_{\max} \to \infty$, which saturates at twice today's detector plateau. All simulated PET trajectories in Figure 4 and Table 4 use Equation (4).[7]

## 3.3 | Residual Detection Floor

Regardless of secrecy and PET improvements, there is always a chance monitoring may uncover clandestine nuclear activities (Hinderstein 2013). We therefore embed floor $\pi_0$ in the logistic,

$$\Pr(D_t) = \pi_0 + \frac{1 - \pi_0}{1 + \exp[\kappa\, \text{RAI}(t) + \theta]}, \qquad \pi_0 = 0.10, \quad (5)$$

so detection never falls below 10%.

## 3.4 | Modeling Opportunistic Breakout

State nuclear breakout attempts are complex political decisions driven by security concerns, domestic politics, and status considerations (Sagan 1996). A comprehensive model of proliferation would need to account for relevant geopolitical factors. The purpose of our model, however, is to isolate the effect of technological change on proliferation risk. We therefore simplify the drivers of breakout attempts into a hazard rate, $\lambda(t)$, representing the background rate at which states might be tempted to pursue clandestine weaponization. This approach allows us to analyze how technological balance—captured by RAI—might influence this rate, holding constant deeper political drivers of proliferation.

We acknowledge that significantly positive RAIs may encourage latent proliferators to attempt breakout. Thus, we elevate baseline hazard $\lambda_0 = 0.05$ year$^{-1}$ by an "opportunism" boost increasing alongside RAI. We denote by $\tau$ the threshold above which proliferators judge detection risk sufficiently low and opportunistic breakout incentives materialize. Because small RAI changes are difficult for states to evaluate, we delay incentives for clandestine weaponization until a gap favoring PET is unambiguous. The boost term accordingly uses a shifted sigmoid curve,

$$\lambda(t) = \lambda_0\left[1 + \eta\, \sigma(\beta[\text{RAI}(t) - \tau])\right], \quad \sigma(x) = \frac{1}{1 + e^{-x}}, \quad (6)$$

with $\tau = 40$ and $\beta = 0.10$. Here, $\tau = 40$ represents a significant, sustained proliferator advantage, hypothesized as a threshold wherein perceived detection risk substantially diminishes. The maximum hazard multiplier related to $\eta$ (where baseline $\eta = 3$ implies a fourfold increase in $\lambda(t)$ if $\sigma(\cdot) \to 1$) is conceptualized to capture a scenario where two to three states, previously deterred, might initiate covert programs due to advanced PETs and lagging DETs. While these parameters are based on plausibility rather than precedent, their significant influence (see Appendix A) underscores the importance of understanding and mitigating opportunistic proliferation. The hazard doubles only when the proliferator is $\approx 45$ points ahead and approaches its fourfold ceiling once RAI > 80.

Equation (1) becomes

$$R(T) = \int_0^T \lambda(\text{RAI}(t))\left[1 - \Pr(D_t)\right] dt, \quad (7)$$

where $\Pr(D_t)$ is given by logistic form (5) (reduced to Equation 2) with $\kappa = 0.4$ under status quo detectors and $\kappa = 0.6$ once the first part of a "moonshot" detection technology package is deployed in year 1.

Equation (6) describes opportunistic breakout hazard rate $\lambda(t)$, a function of $RAI(t)$ increasing sigmoidally once $RAI(t)$ surpasses threshold $\tau$. Figure 3 depicts this relationship, showing how breakout attempt rate (hazard) escalates as perceived proliferator advantages grow. For instance, we may state a simple condition:

**Lemma 1.** *If $RAI(t) \gg \tau$ for a substantial portion of period $[0, T]$, such that $\sigma(\beta[RAI(t) - \tau]) \approx 1$, and given $\pi_0$ is small, cumulative risk $R(T)$ will be driven significantly by the boosted hazard, approaching $\lambda_0(1 + \eta)T\, \mathbb{E}[1 - Pr(D_t)]$.*

This proposition highlights that sustained high RAI not only increases per-attempt evasion probability but also the attempt rate itself.[8]

 

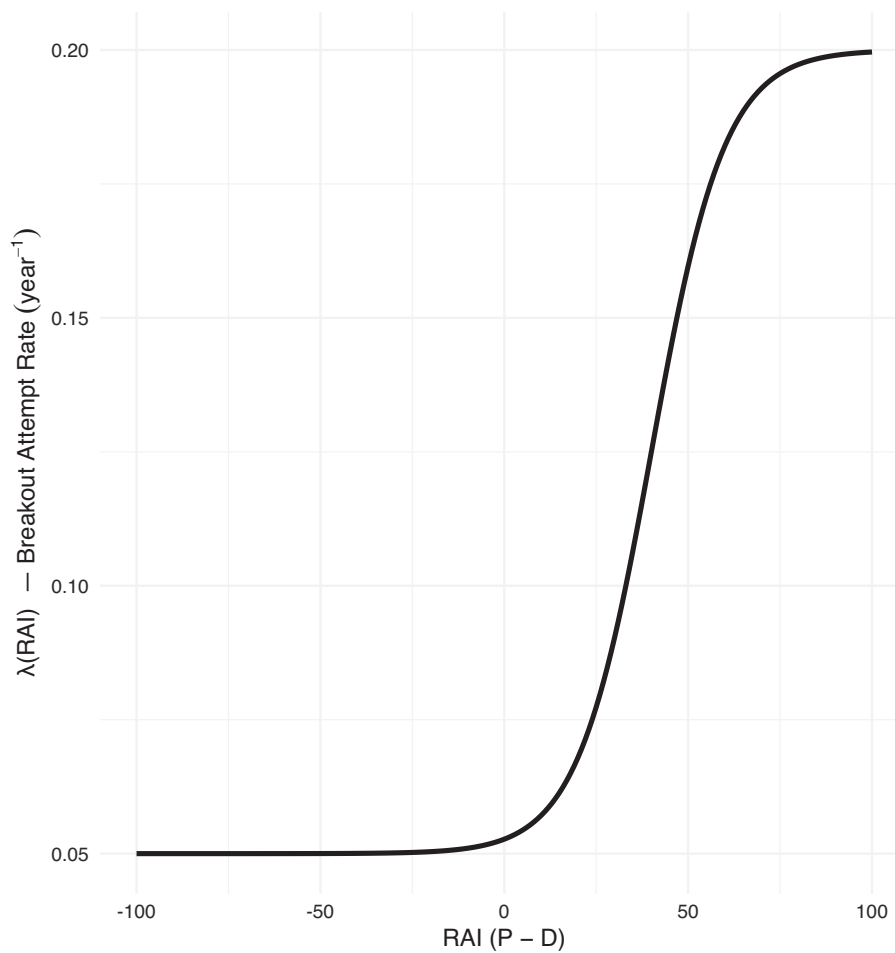

**FIGURE 3** | Opportunistic covert nuclear proliferation breakout attempt rate.

**TABLE 1** | Parameters.

| Symbol | Meaning |
| --- | --- |
| $P(t)$ | Proliferation-enabling capability index at time $t$ |
| $D(t)$ | Detection-enhancing capability index at time $t$ |
| $RAI(t)$ | Relative Advantage Index, defined as $P(t) - D(t)$ |
| $P_0$ | Initial value of $P(t)$ at $t = 0$ (2025 baseline) |
| $D_0$ | Initial value of $D(t)$ at $t = 0$ (2025 baseline) |
| $g_P (year^{-1})$ | Logistic growth coefficient for PET |
| $P_{\max}$ | Logistic ceiling (carrying capacity) for $P(t)$ |
| $\kappa$ | Logistic slope parameter in detection probability mapping |
| $\theta$ | Logistic intercept (shift) parameter in detection probability mapping |
| $\pi_0$ | Residual (minimum) detection probability floor |
| $\lambda_0$ | Baseline annual breakout-attempt hazard rate |
| $\eta$ | Maximum proportional increase in opportunistic breakout hazard |
| $\beta$ | Logistic slope governing sensitivity of opportunistic hazard to $RAI(t)$ |
| $\tau$ | $RAI(t)$ threshold above which opportunistic breakout hazard increases |
| $R(T)$ | Expected number of undetected breakout attempts over time horizon $T$ |
| $P(T)$ | Probability of at least one undetected breakout over time horizon $T$ |

**TABLE 2** | PET improvements $g_P$(year$^{-1}$).

| Scenario | $T_d$ | $g_P$(year$^{-1}$) | Interpretation |
| --- | --- | --- | --- |
| Limited AI | 36 months | 0.23 | Matches diffusion of 3D printing, commercial satellite imagery |
| Disruptive AI | 25 months | 0.33 | "Moore's Law," GPT-4 (2023) to o3 (2025) |
| Transformative AI | 7 months | 1.189 | "Ray's Law" 2025 AI–frontier doubling cadence |

**TABLE 3** | Scenario assumptions.

| Parameter | Limited | Disruptive | Transformative | DET moonshot |
| --- | --- | --- | --- | --- |
| PET growth rate $g_P$(year$^{-1}$) | 0.23 | 0.33 | 1.19 | 0.23, 0.33, 1.19 |
| Step gains in DET Δ | +5, +4 | +5, +4 | +5, +4 | +12, +10, +10 |
| Logistic slope $\kappa$ | 0.4 | 0.4 | 0.4 | 0.6 |

*Note:* $P_{max} = 120$, $\pi_0 = 0.10$.

## 3.5 | Interpretation

Ceiling (4) and floor (5) together capture diminishing marginal returns of ever-larger language models. Design tasks compress quickly, but material throughput and physics impose hard stops. Opportunistic hazard (6) then shows how, even under those physical limits, risk climbs steeply if the international community delays "moonshot" detector upgrades beyond year ∼ 3. Table 1 recaps the model parameters.

## 4 | Six Illustrative Scenarios

To illustrate effects of technological and policy changes on nuclear proliferation risk, we present six scenarios. They reflect intersections of three paths for PET and two for DET. In all tables and figures, $R(10)$ is the expected number of undetected breakout attempts in 10 years. We use the Poisson process as a standard and tractable modeling assumption for events occurring at a given rate over time, acknowledging that real-world nuclear proliferation decisions are not truly independent events. If $R(10) = 0.40$, the system averages 0.4 successful covert breakout attempts per decade. The probability of at least one undetected clandestine weaponization program is

$$P(10) = 1 - e^{-R(10)}.$$

Thus, $R(10) = 0.40$ implies $P(10) = 0.33$; values in our tables use this transformation.

We examine a range of scenarios because the future of AI is subject to intense debate and uncertainty. AI skeptics argue that LLMs cannot generate novel scientific knowledge because they are trained on existing scientific data and limited by the current scientific frontier. But this perspective is weakening: Recent research shows large models can generate novel proofs, reaction mechanisms, and algorithmic shortcuts (Gibney 2025; Gottweis et al. 2025). Recent history suggests that progress often surprises even the best-informed experts, with general models scaled on massive compute frequently achieving benchmarks years before predicted. Once models produce original scientific insights, the nature of training data limitations may change, allowing for a very high capability ceiling before advancement plateaus. At the same time, it is plausible that current progress will soon hit fundamental data or algorithmic limits. To manage this uncertainty, our analysis does not commit to a single future. Instead, we explore three distinct PET growth trajectories: plateauing capability (Limited AI), steady improvement (Disruptive AI), and massive acceleration (Transformative AI). This allows us to assess policy interventions across several plausible worlds.

We also model two DET paths. The baseline assumes detection improves similarly to recent decades, with two medium-size upgrades of Δ = +5 and +4 (analogous to environmental swipe sampling and routine high-resolution satellite imagery). We set $\kappa = 0.4; \theta = 0$.[9] We compare this to a "moonshot" DET program involving three larger upgrades (Δ = +12, +10, +10) at years 1, 3, and 6, and a steeper slope $\kappa = 0.6$ (e.g., improved sensor fusion, lower false positives).

The first scenario, our baseline, assumes 2000–2020 PET and DET trends persist over the decade. Growth coefficient $g_P = 0.23$ year$^{-1}$ implies enabling technologies (additive manufacturing, CNC machining, others) double every 3 years, mirroring recent tempo (European Patent Office 2023). In this "Limited AI" scenario, generative models plateau at levels akin to capable graduate students.[10] Model weights diffuse widely, but quality gains are incremental and costly. We solve this illustrative case in closed form before proceeding to other simulations.

### 4.1 | Walk-Through: Baseline Limited AI → "Moonshot" Transition

*Step 1: PET Path*

For Limited AI, we use growth coefficient $g_P = 0.23$ year$^{-1}$ (doubling time of approximately 3 years). With $P_0 = 20$ and $P_{max} = 120$, Equation (4) gives

$$P_{\text{base}}(t) = \frac{120}{1 + 5e^{-0.23t}}.$$

*Step 2: DET Path*

Detectors start at $D_0 = 50$ and jump by $\Delta_1 = +5$ at $t = 3$ and $\Delta_2 = +4$ at $t = 7$:

$$D_{\text{base}}(t) = \begin{cases} 50, & 0 \leq t < 3, \\ 55, & 3 \leq t < 7, \\ 59, & t \geq 7. \end{cases}$$

*Step 3: RAI Trajectory*

Setting $P_{\text{base}}(t) = D_{\text{base}}(t)$ and solving

$$\frac{120}{1 + 5e^{-0.23t}} = 55 \implies e^{-0.23t} = \frac{\frac{120}{55} - 1}{5} \implies t^\star = \frac{1}{0.23} \ln\left(\frac{\frac{120}{55} - 1}{5}\right) \approx 6.27 \text{ year.}$$

Because $t^\star$ falls in the [3, 7]-year segment, detectors are ahead until year ∼6.27; thereafter RAI > 0.

*Step 4: Risk Integral Sans "Moonshot"*

Without opportunism ($\eta = 0$) and $\pi_0 = 0.10$, integration of Equation (1) yields

$$R(10)^{\text{base}} = 0.17 \implies P(10)^{\text{base}} = 0.16.$$

*Step 5: Inject DET "Moonshot"*

Three detector upgrades now occur at $\{1, 3, 6\}$ with $\Delta = \{+12, +10, +10\}$ and $\kappa = 0.6$.

$$D_{\text{MS}}(t) = \begin{cases} 50, & 0 \leq t < 1, \\ 62, & 1 \leq t < 3, \\ 72, & 3 \leq t < 6, \\ 82, & t \geq 6. \end{cases}$$

Under "moonshot" DET path $t^\star \to \infty$, RAI stays negative and detectors remain ahead for the decade.

*Step 6: Risk with DET "Moonshot"*

Under the same $\lambda_0 = 0.05$ and $\pi_0$,

$$R(10)^{\text{MS}} = 0.003 \implies P(10)^{\text{MS}} = 0.003.$$

*Assessment*

Early, large monitoring and verification detector jumps cut decade-wide nuclear breakout risk by $\Delta R \approx -98\%$ relative to Limited AI. Subsequent sections show this result erodes as PET growth accelerates or opportunistic weaponization incentives intensify.

## 4.2 | Modeling Continued AI Improvements

Our second scenario reflects HPC influence on PET and LLM improvements from 2020 to 2025. This Disruptive AI scenario features a $g_P$ of 0.33 year$^{-1}$, reflecting the 2-year Moore's Law for transistor doubling (Mack 2011) and the 25-month period between GPT-4-class models (March 2023) and the release of o3 (April 2025). This scenario envisions steady commercial progress, where each foundation model brings notable coding, simulation, and design gains, though returns still diminish as the training corpus is exhausted. Nonproliferation inspection budgets and doctrines remain unchanged, leaving detector gains and the logistic slope unchanged.

The third scenario builds on the second but assumes continued rapid, near-exponential growth of LLM capabilities. This "AI explosion" drives PETs toward their physical ceiling at an accelerated pace. We set $g_P = 1.19$ year$^{-1}$, reflecting "Ray's Law," an AI-specific update to Moore's Law noting some capabilities are doubling every 7 months. This scenario represents the upper bound of current uncertainty. Self-reinforcing advances—agents improving successive models and automated experimenters expanding training sets—sustain or even accelerate this pace, with PET racing to physical ceilings by year 3. Table 2 shows doubling rates for each scenario.

Our fourth, fifth, and six scenarios consider "moonshot" DET investments. "Moonshots" assume the international community quickly invests in three major detector upgrades (e.g., AI-enhanced monitoring, thermal–infrared constellations, or global active–neutral particle sensors), yielding larger stepwise improvements of $\Delta = +12, +10$, and $+10$. Improved sensor fusion steepens the logistic slope to $\kappa = 0.6$. We calculate the effect of "moonshot" interventions on $R(10)$ for the first three scenarios.

We set $\lambda$(year$^{-1}$), the annual probability of a covert weaponization attempt, at 0.05.[11] This is conservative but defensible: First, it reproduces IAEA assessments of ≈90% detection probability. Second, roughly 10 technologically advanced NPT states possess complete fuel-cycle knowledge but have not weaponized (Herzog 2020). Assigning each a modest 0.5% annual probability of initiating a secret program aggregates to $\lambda = 0.005 \times 10 = 0.05$ year$^{-1}$. Finally, the international community has uncovered four clear clandestine nuclear efforts beyond basic research since 1991: Iraq's pre-1991 electromagnetic isotope separation, Iran's undeclared enrichment and heavy water facilities revealed in 2002, Syria's al-Kibar reactor destroyed by Israel in 2007, and the restarted North Korean program after its 1994–2002 freeze (Findlay 2015). Counting each discovery as a "draw" from a Poisson process over the 34-year interval of 1991–2024 gives an empirical hazard $\lambda = 0.11$ year$^{-1}$. However, two cases (Iraq and North Korea) originated before 1991; excluding legacy programs yields $\lambda = 0.06$ year$^{-1}$, close to our assumption of 0.05. Inserting the $g_P$ values from Table 2 into Equations (1)–(2), and keeping the baseline detector upgrades at $\Delta = \{+5, +4\}$ with $\kappa = 0.4$, $\theta = 0$, yields:[12]

$$\begin{aligned} R_0(10) &= 0.17 \quad \text{(Limited AI)} \\ R_1(10) &= 0.26 \quad \text{(Disruptive AI)} \\ R_2(10) &= 0.40 \quad \text{(Transformative AI)} \end{aligned}$$

### 4.3 | Modeling DET "Moonshot"

We now retain the same $g_P$ values but add the "moonshot" DET package. This entails significant investment and policy effort resulting in three large detection technology improvements ($\Delta$ = {+12, +10, +10}) at years 1, 3, and 6 instead of two moderate improvements ($\Delta$ = {+5, +4}) at years 3 and 7, as well as a steeper logistic slope $\kappa$ = 0.6 (e.g., better sensor fusion, lower false-positive rates). Recalculated risks are:

$$R_0^\star(10) = 0.00 \quad \text{(Limited AI } g_0\text{)}$$
$$R_1^\star(10) = 0.13 \quad \text{(Disruptive AI } g_1\text{)}$$
$$R_2^\star(10) = 0.39 \quad \text{(Transformative AI } g_2\text{)}$$

### 4.4 | "Moonshot" Insights

Reducing the doubling interval from 3 years to just over 2 years raises 10-year breakout risk by roughly 50% (0.17 → 0.26). An extreme 7-month cadence more than doubles it. However, a realistic DET "moonshot" nearly eliminates near-term proliferation risk at conventional PET growth rates ($R_0^\star$ = 0.00) and significantly reduces it in the Disruptive AI case ($R_1^\star$ = 0.13). Yet, it fails to contain the Transformative AI scenario: When PET doubles every 7 months, proliferator gains dominate, and the "moonshot" has little impact ($R_2^\star \approx$ baseline). Finally, simulations reveal steeply diminishing returns to detector investment once $g_P >$ 1 year$^{-1}$, suggesting growth governance (e.g., export controls or model-weight constraints) or addressing demand-side incentives may be the only viable paths to contain risk in a Transformative AI world. Table 3 recaps scenario conditions; Figure 4 shows PET, DET, and RAI trajectories.

### 4.5 | Opportunistic Breakout Attempts

Our prior models assumed RAI changes did not affect proliferation likelihood in year $\lambda$year$^{-1}$. Since proliferators may be more likely to attempt covert development if DET significantly lags behind PET, we model opportunistic breakout. We retain baseline exogenous hazard $\lambda_0$ = 0.05 year$^{-1}$, but let it scale up to four times its baseline once the technological balance decisively favors proliferators. This captures the idea that two or three additional latent states may choose to secretly pursue the bomb once advanced AI tools erode tacit knowledge barriers and lagging detection fosters fears of rival proliferation. Table 4 recomputes 10-year risk assessments using updated hazard function equation (7) for the "opportunistic" and "both" conditions.

When RAI($t$) crosses the $\tau$ = 40 threshold, hazard quickly rises toward ceiling 0.20 year$^{-1}$. Afterward, cumulative risk grows almost linearly. With no "moonshot," the Transformative AI variant more than triples to $R(10)$ = 1.44 $(P(10) = 0.76)$, while opportunism in the Disruptive AI case increases risk by a little more than one-third $(R(10) = 0.26 \to R(10) = 0.36)$, driven by positive RAI in years 7–10. Opportunistic proliferation incentives play no role in the Limited AI scenario, and risk remains $R(10) \approx$ 0.17. Opportunism also impacts the effectiveness of the DET "moonshot." It still wipes out almost all risk in the Limited AI scenario (a nearly 100% reduction relative to opportunistic baseline). For the Disruptive AI scenario, the "moonshot" reduces risk by $\approx$64% (from $R(10)$ = 0.36 to $R(10)$ = 0.13) when opportunism is present. Even in the case of Transformative AI with opportunism, "moonshot" detector investments reduce risks by $\approx$37.5% (from $R(10)$ = 1.44 to $R(10)$ = 0.90), though this leaves an unacceptably high undetected proliferation risk. Once PET growth exceeds $g_P \approx$ 1 year$^{-1}$, risk reduction through export controls, model-weight governance, supply-chain policing, or reducing demand-side proliferation incentives offers greater payoffs than detector investments.

### 4.6 | Scenario Insights

The scenarios highlight three messages. First, breakout risk compounds more quickly than intuition suggests. A 43% increase in proliferation growth rate (Limited AI to Disruptive AI), without detector upgrades, increases 10-year probability of undetected proliferation by nearly 90% (from $P(10)$ = 0.16 to $P(10)$ = 0.30), including opportunistic incentives.

Second, early "moonshot" investments in DETs can more than offset even an AI-driven acceleration of PETs. Their discounted cost is modest relative to losses they avert. When the growth curve is steep, moving an improvement in DET forward by 2 years buys more risk reduction than does doubling its magnitude. Figure 5 depicts risk changes over time.

Third, risk elasticity $\varepsilon_R$ is highest when RAI hovers near zero with a steep logistic slope. Each marginal detector improvement yields disproportionately large risk reduction in regimes that matter for policy. This reflects when each scenario crosses the opportunism threshold $\tau$: Limited AI never exceeds $\tau$, so the hazard stays constant. Disruptive AI crosses late (years 8–10), prompting modest risk increases. Transformative AI crosses early (within 2 years), sharply boosting opportunistic proliferation. These patterns strongly reinforce the case for front-loaded verification investments.

Fourth, our model of opportunistic breakout is a deliberate simplification of a complex political phenomenon. The hazard rate, $\lambda(t)$, abstracts from specific, crucial geopolitical drivers, such as the stability of international alliances, regional rivalries, or domestic political pressures for proliferation. Real-world breakout decisions are not independent events and are highly contingent on this political context. A valuable area for future research would be to integrate our model of the technology race into more sophisticated game-theoretic or agent-based models that capture these strategic interactions.

### 4.7 | Limitations and Parameter Sensitivity

Three caveats qualify our simulations. First, PET ceiling $P_{\max}$ = 120 is a stylized proxy for material-throughput constraints. Raising the cap to 140 postpones saturation by 2 years and increases $R(10)$ by roughly 15% in Transformative AI variants, but it leaves Limited AI outcomes nearly unchanged. Second, detection floor $\pi_0$ = 0.10 is arguably pessimistic. Raising it to $\pi_0$ = 0.15 halves $R(10)$ in the explosion cases, yet barely moves the baseline, highlighting the value of improving residual signature sampling. Third, while the opportunistic-hazard boost ($\eta = 3$, $\beta = 0.10$) reflects historical clandestine program frequency, doubling $\eta$ or halving $\beta$ renders the "both" scenarios risk-dominant in under 5 years. This suggests governance measures dampening proliferation incentives can be as valuable as detector "moonshots" once PET growth accelerates. Finally, the breadth of possible technological advances and policy choices introduces significant uncertainty. Figure 4 visualizes this; widening bands reflect parameter spreads for timing ($\tau$) and magnitude ($\Delta$) of PET and DET improvements. "Moonshot" detection investments mitigate moderate AI growth proliferation risks, while growth governance becomes essential in a Transformative AI world.

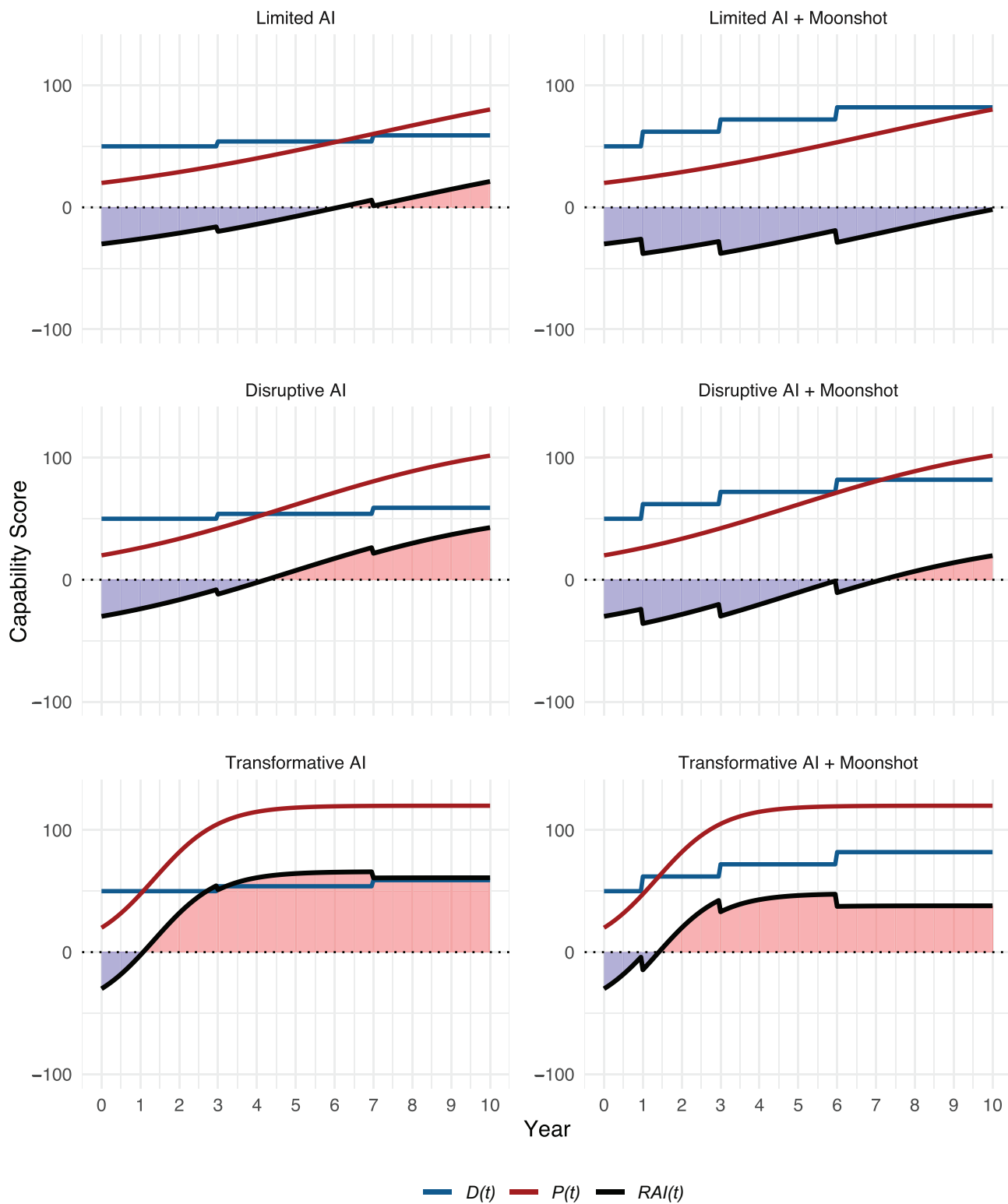

**FIGURE 4** | Six technology race scenarios.

A sensitivity sweep shows 10-year breakout risk depends on opportunistic breakout boost $\eta$ and residual detection floor $\pi_0$.[13]

**TABLE 4** | 10-year risk and probability changes across scenarios.

| Scenario | R(10) | P(10) | ΔR(%) | ΔP(%) |
| --- | --- | --- | --- | --- |
| Limited AI | 0.17 | 0.16 | — | — |
| Limited AI + "Moonshot" | 0.00 | 0.00 | −98 | −98 |
| Limited AI + Opportunistic | 0.17 | 0.16 | 0 | 0 |
| Limited AI + Both | 0.00 | 0.00 | −98 | −98 |
| Disruptive AI | 0.26 | 0.23 | — | — |
| Disruptive AI + "Moonshot" | 0.13 | 0.12 | −49 | −46 |
| Disruptive AI + Opportunistic | 0.36 | 0.30 | 38 | 32 |
| Disruptive AI + Both | 0.13 | 0.12 | −49 | −46 |
| Transformative AI | 0.40 | 0.33 | — | — |
| Transformative AI + "Moonshot" | 0.39 | 0.32 | −2.5 | −4 |
| Transformative AI + Opportunistic | 1.44 | 0.76 | 258 | 130 |
| Transformative AI + Both | 0.90 | 0.59 | 123 | 79 |

*Note:* Percentage changes (ΔR(%), ΔP(%)) calculated relative to baseline specific to each PET growth path (Limited AI, Disruptive AI, or Transformative AI). Baseline is always the respective PET scenario assuming no "moonshot" DET investment and no "opportunistic" breakout behavior. For example, changes for "Limited AI + Moonshot" are relative to the "Limited AI" (no moonshot, no opportunism) scenario.

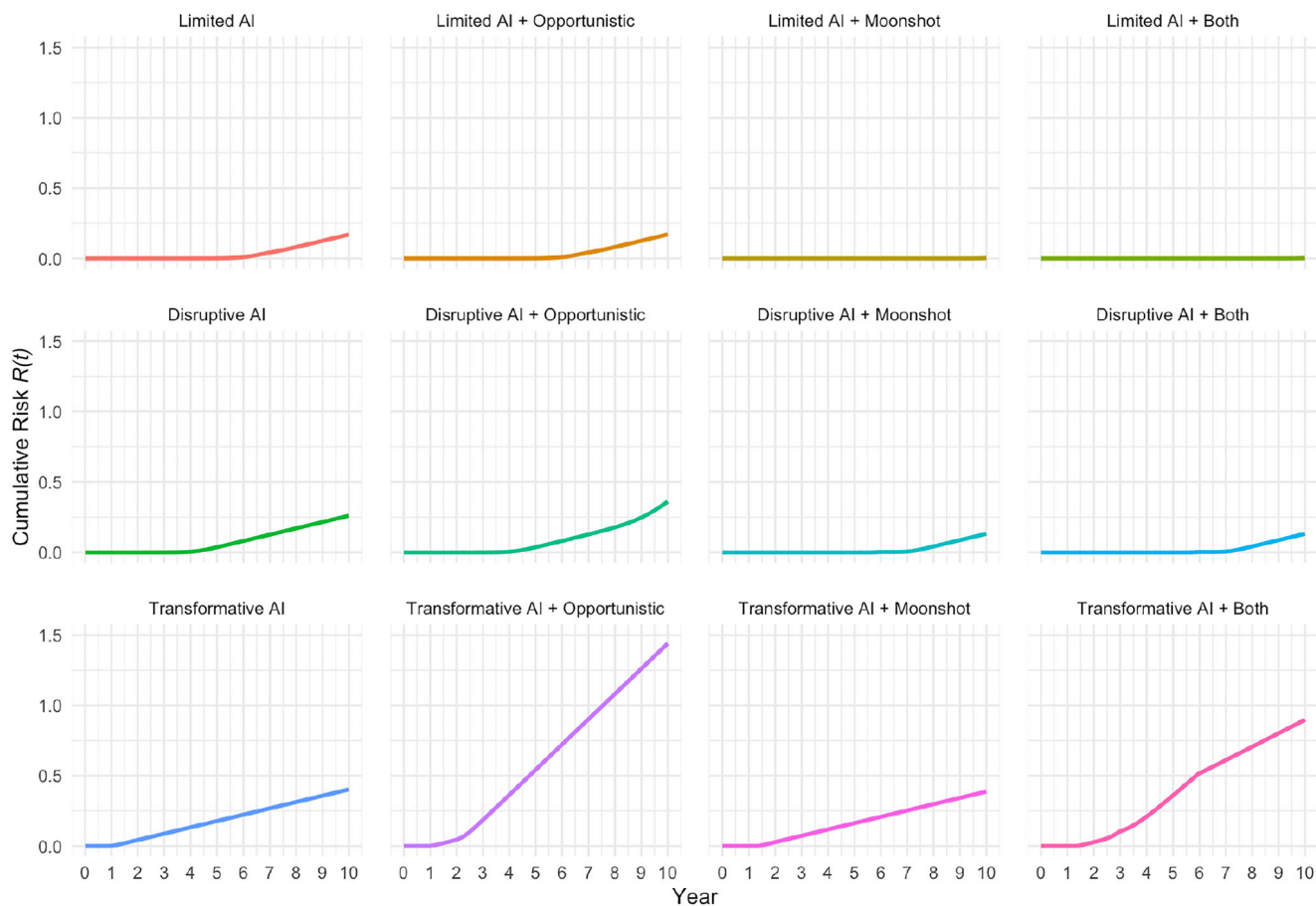

**FIGURE 5** | Cumulative risk $R(t)$.

Increasing $\eta$ from 1 to 5 more than doubles $R(10)$ across PET growth paths, whereas lowering $\pi_0$ from 0.25 to 0.05 raises $R(10)$ by roughly one-quarter, removing a critical backstop in the detection probability curve. In contrast, logistic ceiling $P_{max}$ and opportunistic threshold $\tau$ exert second-order effects, shifting $R(10)$ by about 40% over their full ranges. Detection slope $\kappa$ and hazard slope $\beta$ have minimal effect. Policymakers should note that raising the detection floor (e.g., with wider global radioisotope-ratio sampling) and reducing breakout opportunism (e.g., with tailored security assurances) yield the greatest marginal risk reductions. Further sharpening detector specificity, or making small adjustments to how quickly a state can move

 

from capability to breakout (sprint responsiveness), provides little additional benefit.

Finally, to further enhance accessibility and transparency, we have developed an interactive browser-based simulation tool of our model.[14] The tool allows users to explore how PET and DET shifts influence the RAI and cumulative proliferation risk. Users can toggle between preset scenarios and adjust core model parameters—such as PET doubling time, detection thresholds, and initial conditions—to assess how the model behaves under different assumptions. This feature is especially useful for those who may question our baseline inputs, as it allows for visualization of the consequences of alternative assumptions.

## 5 | Conclusion

Confronted with an asymmetric technology race and deepening uncertainty, the international community must adapt nonproliferation risk management approaches. A proactive and adaptive posture is required to shift the dynamics of the race or at least mitigate the gravest risks that may lead more states to acquire the bomb. Verification has always been a "club good," wherein a handful of states finance costly safeguards benefiting all (Hunt 2022). LLMs and other emerging PETs invert the economics. General-purpose models compress design timelines and diffuse tacit knowledge at consumer electronics prices, while detector upgrades remain capital intensive and built-for-purpose (International Atomic Energy Agency 2022; Muñoz et al. 2024; Özcan 2024; Stewart & Mayhew 2019).

To prevent detection from falling irreparably behind, rapid DET innovation is essential. Our simulations (Table 4) show "moonshot" investments can substantially reduce breakout risks, especially under moderate PET growth. This could take the form of a Frontier Detection Acceleration Initiative (FDAI)—a dedicated effort involving governments and their NTM, international monitoring and verification organizations like the IAEA and CTBTO, and the commercial tech sector. This last component is essential. As the PET frontier is driven by companies (Herzog & Kunertova 2024; Vaynman & Volpe 2023), viable arms control must integrate the private sector not only as a regulated actor but as a partner in verification and governance innovation.

The FDAI could include innovation prizes or grants for AI-enabled inspection tools, collaborative public–private partnerships to prototype new sensors, and mechanisms to deploy promising technologies without bureaucratic delays. Red-teaming emerging proliferation pathways would be institutionalized, with results feeding rapid-response upgrades. The goal would be to reduce gaps between PET advances and DET deployment, moving from reactive to anticipatory posture.

The RAI highlights how PET–DET dynamics reshape strategic incentives. When PETs significantly outpace DETs (high positive RAI), states may see clandestine nuclear proliferation as more viable and lower risk. Conversely, strong and improving detection (negative RAI) may steer states toward hedging or restraint. Shifts in RAI, especially if sudden or unanticipated, could cause leaders to miscalculate windows of vulnerability or opportunity. Our model isolates the relevant technological factors, but future research should endeavor to more formally integrate RAI shifts into complex political models of proliferation.

The institutional rigidity of current agreements compounds the pacing problem (Marchant 2011). Treaties institutionalize verification that may become obsolete, particularly when dealing with rapid technological development (Canfil 2025b). Agile verification regimes are essential, encompassing adaptive, forward-looking clauses or standing technical bodies empowered to revise measures as technologies evolve (Canfil 2025a; Egel 2025). For instance, the IAEA could regularly review the need to integrate emerging techniques like AI-driven anomaly detection or wide-area environmental sampling into its safeguards implementation under the Additional Protocol. It could likewise make DET development requests through Member State Support Programmes, based upon careful analysis of AI trajectories and other technology trends. Modular coalitions of volunteer states could pilot these next-generation tools and feed lessons into global norms. The reality is that to contain tomorrow's nuclear risks, monitoring and verification bodies must operate more like startups than bureaucracies: iterative and experimental.

Safeguards thinking must become a part of emerging technology governance. Many PETs—AI, additive manufacturing, synthetic biology—are dual-use but governed by fragmented or nonexistent regimes. For AI, no dedicated international framework exists (Brockmann 2022). Some AI developers test for dangerous capabilities or restrict open weights, but export control regimes like the Wassenaar Arrangement must be updated to cover open-weight foundation models that could be fine-tuned for proliferation. Embedding "safeguards-by-design" into frontier systems—like watermarking AI outputs or embedding auditability in manufacturing devices—can improve verification. New hybrid institutions at the intersection of national security, AI safety, and scientific integrity may likewise be necessary.

On the demand side, geopolitical tensions and security concerns remain the most potent driver of nuclear weapons pursuit. Our elasticity results (Table 5) show a clear strategic fork: Moderate PET growth can be contained through "moonshot" DET investment; under Transformative AI conditions, this strategy yields diminishing returns (Table 4). At that point, the most powerful interventions likely shift toward robust PET growth governance (export controls, model-weight constraints, or ethical AI use and development norms) and reducing incentives for states to seek nuclear weapons in the first place. Without such measures, traditional detection systems—even upgraded—may be overwhelmed.

Importantly, none of this is deterministic. The future of AI, the willingness of states to proliferate, and the evolution of multilateral cooperation are contingent. But the asymmetry of logistic PET scaling versus stepwise DET improvement means delays could prove dangerous. Even modest PET advances can compound into strategic gaps. This creates urgency for investments in next-generation arms control, monitoring, and verification. Indeed, the ongoing crisis over Iran's nuclear program is a reminder that nuclear proliferation is hardly a solved problem. With additive manufacturing of centrifuge components, AI-guided cascade optimization, and simulation-driven warhead design, proliferators may be able to accelerate

TABLE 5 | Risk, probability, and elasticity[a] relative to Disruptive AI baseline ($R(10) = 0.26$).

| Scenario | $R(10)$ | $P(10)$ | $\varepsilon R$ |
| --- | --- | --- | --- |
| Limited AI | 0.170 | 0.157 | −0.348 |
| Limited AI + Opportunistic | 0.171 | 0.157 | −0.345 |
| Limited AI + "Moonshot" | 0.003 | 0.003 | −0.989 |
| Limited AI + Both | 0.003 | 0.003 | −0.989 |
| Disruptive AI | 0.261 | 0.230 | 0.000 |
| Disruptive AI + Opportunistic | 0.361 | 0.303 | 0.383 |
| Disruptive AI + "Moonshot" | 0.132 | 0.124 | −0.494 |
| Disruptive AI + Both | 0.133 | 0.124 | −0.492 |
| Transformative AI | 0.403 | 0.332 | 0.543 |
| Transformative AI + Opportunistic | 1.441 | 0.763 | 4.516 |
| Transformative AI + "Moonshot" | 0.388 | 0.321 | 0.485 |
| Transformative AI + Both | 0.899 | 0.593 | 2.441 |

[a] $\varepsilon_R$ calculated as percentage change in cumulative 10-year risk $R(10)$ resulting from a sustained one-point increase in $RAI(t)$ across the decade, serving as an aggregate sensitivity measure.

nuclear weapons breakout in less predictable and detectable ways.

This challenge is unlike past nuclear risks. PET drivers are now decentralized, private, and global (Rosenbach et al. 2025). These tools are increasingly no longer property of national weapons labs. This necessitates supplementing traditional tools like treaties, export controls, and inspections with adaptive, technologically informed strategies. Ultimately, detection and enforcement must evolve at the speed of technology, becoming as innovative as tools available to would-be proliferators. This may seem unnatural for institutions accustomed to cautious diplomacy and consensus-based processes, but the alternative is falling hopelessly behind.

If the capacity to hide nuclear weapon programs outpaces the ability to find them, the world could enter an era where nonproliferation norms crumble. The race is on, and it will not be won with yesterday's methods.

**Acknowledgments**

The authors are grateful for helpful feedback from *Risk Analysis* area editor Vicki Bier, two anonymous peer reviewers, and a number of technical experts in nuclear weapons design and frontier AI models. Assistance from OpenAI's o3 large-language model was used to check mathematical derivations, debug simulation code, and resolve LaTeX formatting issues. All model outputs were critically reviewed, edited, and approved by the authors, who assume full responsibility for the final manuscript. Stephen Herzog began this research while working at the Center for Security Studies at ETH Zurich.

Open access publishing facilitated by Eidgenossische Technische Hochschule Zurich, as part of the Wiley - Eidgenossische Technische Hochschule Zurich agreement via the Consortium Of Swiss Academic Libraries.

**Conflicts of Interest**

The authors declare no conflicts of interest.

**Endnotes**

[1] See Chiou (2025), Wong (2024), and Yin (2025).

[2] Our simulations are robust to changes in baseline RAI so long as $D(t) > P(t)$.

[3] Some recent evidence suggests LLM capabilities are approaching diminishing returns (Villalobos et al. 2024).

[4] The CTBTO likewise aims to detect 1-kiloton nuclear explosions with 90% success or better (Dahlman et al. 2011).

[5] We assume equal PET and DET footing gives a program a 50% chance of detection. Further parameter tuning and sensitivity analyses appear in Appendix A.

[6] "Superintelligent design aids" denotes highly advanced AI systems functioning as powerful tools for human designers. Our analysis does not address distinct speculative risks associated with emerging autonomous, goal-directed superintelligence. This would represent a different class of risk phenomenon beyond the scope of our model.

[7] Setting $P_{\max}$ higher delays saturation and raises late-decade risk tails in our simulation window.

[8] Appendix B offers a proof. It follows from integrating Equation (6) over the interval where $RAI(t) \gg \tau$ and approximating $\sigma(\beta[RAI(t) - \tau]) \approx 1$.

[9] In Equation (2), logistic slope $\kappa$ gauges how an additional point of RAI moves detection odds, while $\theta$ centers the curve. We use $\kappa = 0.4$ for status quo detectors and $\kappa = 0.6$ once "moonshot" upgrades improve sensor fusion.

[10] At the time of writing, frontier models show high performance on many standardized tests. This capability can be misleading, however, as this does not necessarily equate to human competence. Even the best models do not eliminate the need for human review of complex, high-stakes tasks (Johri et al. 2025).

[11] Where $\lambda$ represents the hazard faced by the system, not any single state.

[12] 10-year cumulative risk $R(10)$ is evaluated with constant breakout attempt rate $\lambda = 0.05$ yr$^{-1}$, initial conditions $P_0 = 20$, $D_0 = 50$, and detector upgrades at years 3 and 7. Results are stable to ±20% changes in $\lambda$ or the initial $P_0/D_0$ gap.

[13] See Appendix A.

[14] The simulation tool is accessible here: https://david-m-allison.github.io/ProliferationSimulation.

## Appendix A: Robustness Checks

We calibrated the model to baseline vector

$$(P_{max}, \pi_0, \eta, \beta, \kappa, \tau) = (120, 0.10, 3, 0.10, 0.40, 40).$$

Sensitivity analysis now varies all six parameters one at a time. For every grid point, the coupled system Equations (1)–(6) is integrated from $t = 0$ to $t = 10$ year and cumulative breakout risk

$$R(10) = \int_0^{10} \lambda(t) \left[1 - \Pr(D_t)\right] dt$$

is calculated for the three proliferation-enabling technology (PET) regimes.

### A.1. | Results

Figures A.1 and A.2 depict two diagnostics. Figure A.1 shows absolute risk surface $R(10)$ under the Transformative Artificial Intelligence (AI) PET path as each parameter is perturbed across its grid. Figure A.2 presents elasticity

$$\varepsilon_R = \frac{R(10; \theta^+) - R(10; \theta^-)}{R(10; \theta^0)} \frac{\theta^0}{\theta^+ - \theta^-},$$

computed between extreme values $\theta^-$ and $\theta^+$. Note that this arc elasticity differs from point elasticity defined in Equation (3), which measures local sensitivity of $R$ to marginal Relative Advantage Index (RAI) changes. The qualitative ranking Transformative AI > Disruptive AI > Limited AI remains unchanged, corroborating our analysis. Figure A.3 plots the detection probability function

$$\Pr(D \mid RAI) = \pi_0 + \frac{1 - \pi_0}{1 + \exp\left[-(\kappa \, RAI + \theta)\right]}$$

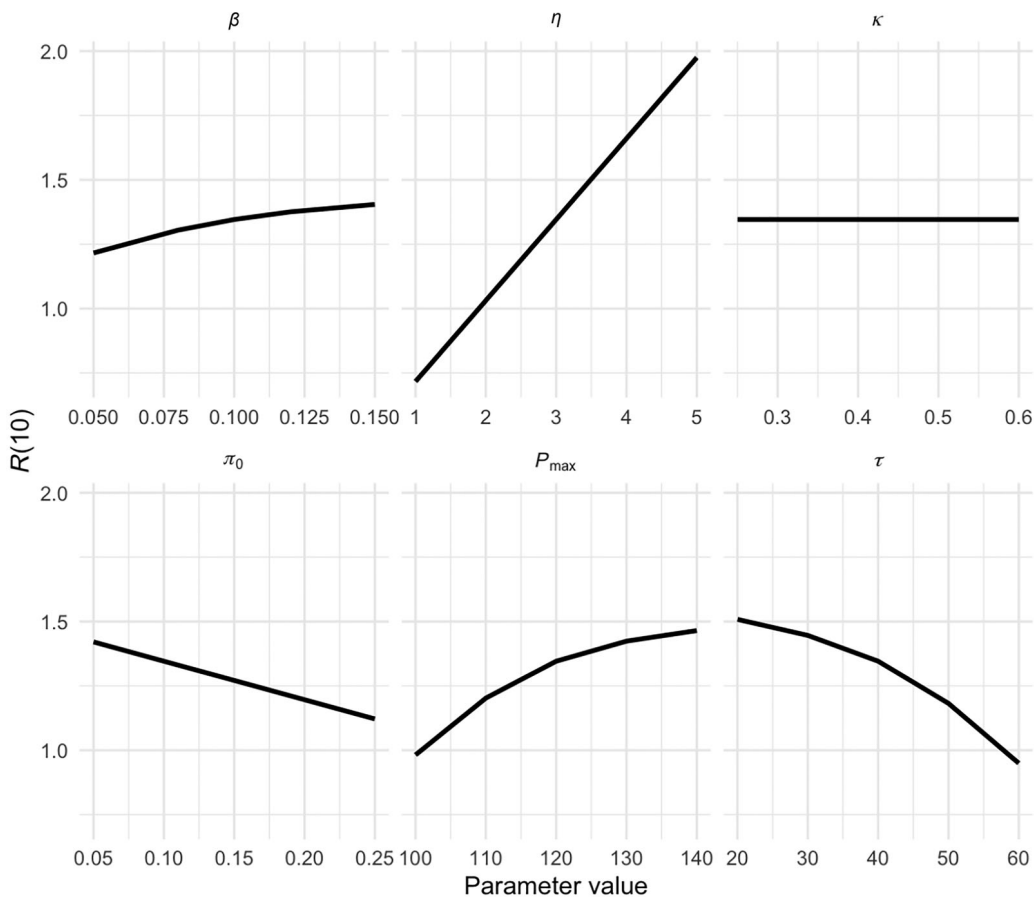

**FIGURE A.1** | $R(10)$ sensitivity to calibrated parameters: Absolute risk for Transformative AI.

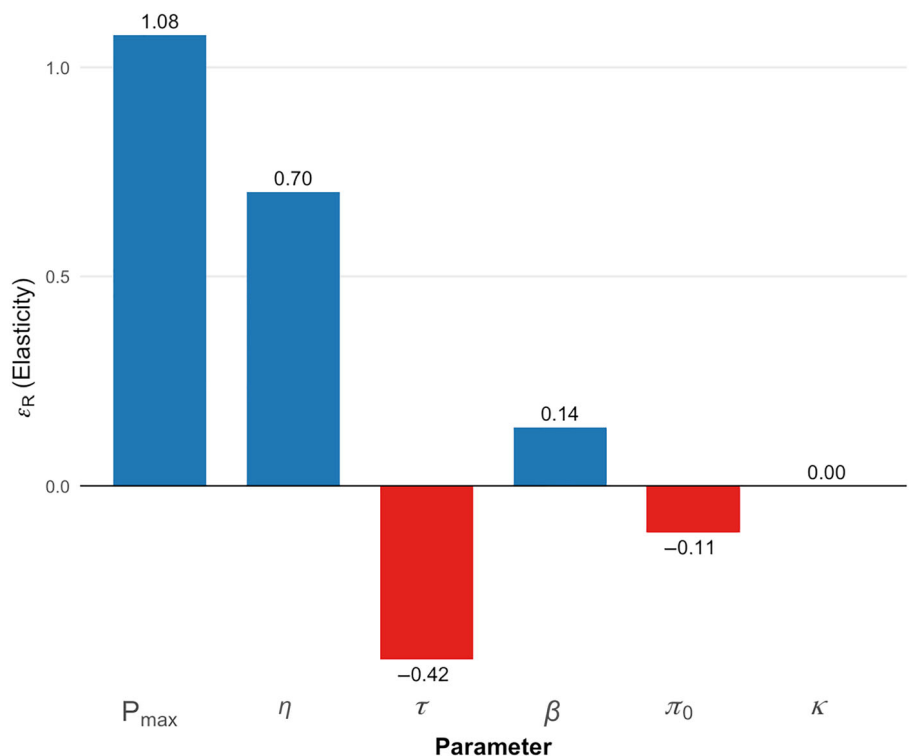

**FIGURE A.2** | $R(10)$ sensitivity to six calibrated parameters: Elasticity $\varepsilon_R$.

**TABLE A.1** | Parameter ranges explored in robustness grid.

| Parameter | Baseline | Range considered |
| --- | --- | --- |
| $P_{max}$ | 120 | 100, 110, 120, 130, 140 |
| $\pi_0$ | 0.10 | 0.05, 0.10, 0.15, 0.20 |
| $\eta$ | 3 | 1, 2, 3, 4, 5 |
| $\beta$ | 0.10 | 0.05, 0.08, 0.10, 0.12, 0.15 |
| $\kappa$ | 0.40 | 0.25, 0.33, 0.40, 0.50, 0.60 |
| $\tau$ | 40 | 20, 30, 40, 50, 60 |

for five values of $\theta \in \{-20, -10, 0, 10, 20\}$, holding $\kappa = 0.4$ and $\pi_0 = 0.10$. The horizontal axis shows RAI $= P(t) - D(t)$, while the vertical axis shows $\Pr(D)$ or detection probability. A dashed horizontal line at $\Pr(D) = 0.50$ marks the logistic inflection point. When $\theta = -20$, the 50% detection threshold occurs at RAI $= 50$; detection remains near 100% until proliferator advantage exceeds 50. In contrast, $\theta = -10$ shifts the 50% threshold to RAI $= 25$, allowing $\Pr(D)$, the detection probability, to become less than certain once $P(t) - D(t)$ surpasses 25. When $\theta = 0$, the threshold lies exactly at RAI $= 0$, so a positive net advantage immediately reduces detection below 50%. For $\theta = 10$ and $\theta = 20$, inflection points lie at RAI $= -25$ and RAI $= -50$, respectively, implying that detectors require a substantial lead to achieve better than 50% efficacy. For Transformative AI scenarios, simulated RAI$(t)$ trajectory exceeds 50 during the middle years, which explains why even $\theta = -20$ fails to guarantee perfect detection (subject to physical limits). Figure A.3 demonstrates that increasingly negative $\theta$ values push the logistic curve so far to the right that detection remains effectively perfect until very high values of RAI, whereas positive $\theta$ values shift the curve left and allow substantial undetected risk whenever proliferators hold a modest lead (Table A.1).

While selecting $\theta = 0$ and thus setting $\Pr(D) = 0.50$ when PET $=$ DET is a logical assumption, future research to determine actual historical and current values of $\theta$ would be a valuable contribution.

### A.2. | Discussion

The strongest policy levers are, in order of effect size, opportunistic breakout boost $\eta$ and residual detection floor $\pi_0$. Increasing $\eta$ from 1 to 5 more than doubles 10-year cumulative breakout risk $R(10)$ across all three PET growth paths. Lowering $\pi_0$ from 0.25 to 0.05—thereby removing the most reliable backstop in the detection-probability curve—raises $R(10)$ by approximately 25%. By contrast, relaxing the logistic ceiling to $P_{max} = 140$, or shifting the opportunistic threshold to $\tau = 60$, changes $R(10)$ by roughly 40%, while detection slope parameter $\beta$ and steepness coefficient $\kappa$ have a negligible impact. These results imply that strengthening the detection floor (e.g., through wider global radioisotope-ratio sampling and multimodal fusion) and dampening opportunistic incentives (e.g., via tailored security assurances and multilateral fuel cycle services) yield the highest marginal reductions in nuclear breakout risk. Incremental improvements in detector specificity, or how opportunistic breakout hazard rises once RAI exceeds threshold $\tau$, offer diminishing returns.

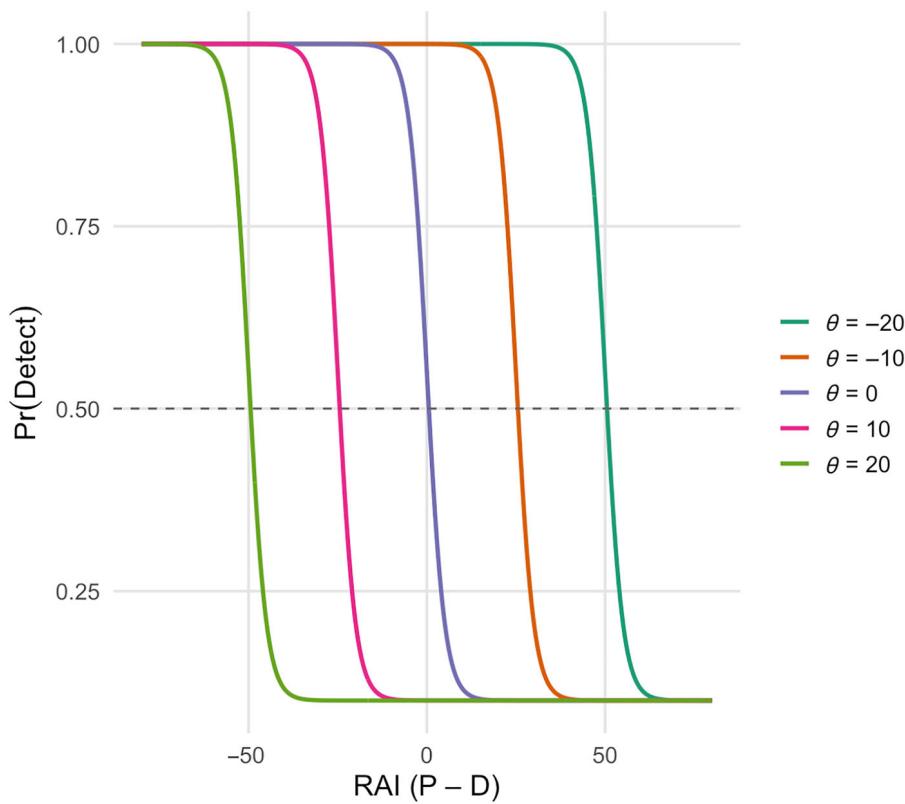

**FIGURE A.3** | Sensitivity of detection to changes in $\theta$.

## Appendix B: Proof of Proposition 1

*Proof.* Let $T > 0$ be the planning horizon. Recall that

$$R(T) = \int_0^T \lambda(t)\left[1 - \Pr(D_t)\right] dt, \qquad \lambda(t) = \lambda_0\left[1 + \eta\,\sigma\bigl(\beta[\text{RAI}(t) - \tau]\bigr)\right],$$

with $\sigma(x) = 1/(1 + e^{-x})$, $\eta > 0$, $\beta > 0$ and threshold $\tau$. Assume that $\text{RAI}(t) \geq \tau + \delta$ on the subinterval $[0, T]$ for some $\delta > 0$. Then,

$$\beta\left[\text{RAI}(t) - \tau\right] \geq \beta\delta > 0 \quad \Longrightarrow \quad \sigma\bigl(\beta[\text{RAI}(t) - \tau]\bigr) \geq \frac{1}{1 + e^{-\beta\delta}} = 1 - \varepsilon,$$

where $\varepsilon := e^{-\beta\delta}/(1 + e^{-\beta\delta}) \leq e^{-\beta\delta}$. Hence,

$$\lambda_0\bigl(1 + \eta(1 - \varepsilon)\bigr) \leq \lambda(t) \leq \lambda_0(1 + \eta). \tag{B.1}$$

Next, $\Pr(D_t)$ obeys

$$\Pr(D_t) = \pi_0 + \frac{1 - \pi_0}{1 + \exp[\kappa\,\text{RAI}(t) + \theta]} \quad \text{with} \quad \kappa > 0.$$

Because $\kappa\,\text{RAI}(t) \geq \kappa(\tau + \delta) \gg 1$, the exponential term in the denominator is at least $e^{\kappa(\tau+\delta)}$. Consequently,

$$\pi_0 \leq \Pr(D_t) \leq \pi_0 + (1 - \pi_0)e^{-\kappa(\tau+\delta)}. \tag{B.2}$$

Define $\zeta := (1 - \pi_0)e^{-\kappa(\tau+\delta)} \ll 1$; then

$$1 - \pi_0 - \zeta \leq 1 - \Pr(D_t) \leq 1 - \pi_0. \tag{B.3}$$

Integrating (B.1) and (B.3) over $[0, T]$ gives the two-sided bound

$$\lambda_0\bigl(1 + \eta(1-\varepsilon)\bigr)\bigl(1 - \pi_0 - \zeta\bigr)T \;\leq\; R(T) \;\leq\; \lambda_0(1+\eta)\bigl(1-\pi_0\bigr)T.$$

Because both $\varepsilon$ and $\zeta$ decay exponentially in $\delta$, letting $\delta \to \infty$ (equivalently, assuming $\mathrm{RAI}(t) \gg \tau$) collapses the lower and upper bounds to a common limit,

$$R(T) \;\longrightarrow\; \lambda_0(1+\eta)\,T\,\mathbb{E}\bigl[1 - \Pr(D_t)\bigr],$$

where the expectation reduces to $1 - \pi_0$ under the same limiting conditions. This establishes the proposition. □

Replicable R scripts are available from the authors, as is a browser-based version of the mathematical model that illustrates its robustness under different parameter specifications: https://david-m-allison.github.io/ProliferationSimulation.